\documentclass[journal=aamick,manuscript=article]{achemso}
\setkeys{acs}{maxauthors = 0}

\usepackage[version=3]{mhchem} 

\usepackage{gensymb}
\usepackage{graphicx}
\usepackage{dcolumn}
\usepackage{bm}
\usepackage{units}
\usepackage{upgreek}
\usepackage{makecell}
\usepackage{xcolor}
\usepackage{soul} 
\usepackage{hyperref}

\newcommand{\ie}{{\it i.e.}, }
\newcommand{\ramcm}{cm$^{-1}$}
\newcommand{\ramcmspace}{cm$^{-1}$ }

\DeclareRobustCommand{\citer}[1]{%
  \begingroup
    \romannumeral-`\x 
    \setcitestyle{numbers}
    \cite{#1}%
  \endgroup   
}

\author{Johannes M. A. Lechner}
\author{Pietro Marabotti} 
\affiliation{Institut f\"ur Physik und IRIS Adlershof, Humboldt Universit\"at zu Berlin, Berlin, Germany}
\author{Lei Shi}
\affiliation{School of Materials Science and Engineering, Sun Yat-sen University, Guangzhou, China}
\author{Thomas Pichler}
\affiliation{Fakultät für Physik, Universität Wien, Wien, Austria}
\author{Carlo Spartaco Casari}
\affiliation{Dipartimento di Energia, Politecnico di Milano, Milano, Italy}
\author{Sebastian Heeg}
\affiliation{Institut f\"ur Physik und IRIS Adlershof, Humboldt Universit\"at zu Berlin, Berlin, Germany}
\email{sebastian.heeg@physik.hu-berlin.de}

\title{Universal Vibrational Anharmonicity in Carbyne-like Materials}

\begin{document}



\begin{abstract}

Carbyne, an infinite linear chain of carbon atoms, is the truly one-dimensional allotrope of carbon. While ideal carbyne and its fundamental properties have remained elusive, carbyne-like materials like carbyne chains confined inside carbon nanotubes are available for study. Here, we probe the longitudinal optical phonon (C-mode) of confined carbyne chains by Raman spectroscopy up to the third overtone. We observe a strong vibrational anharmonicity that increases with decreasing C-mode frequency, reaching up to $8\%$ for the third overtone. Moreover, we find that the relation between vibrational anharmonicity and C-mode frequency is universal to carbyne-like materials, including ideal carbyne. This establishes experimentally that carbyne and related materials have pronounced anharmonic potential landscapes which must be included in the theoretical description of their structure and properties.
\end{abstract}

\newpage
\section{Introduction}\label{SEC:Intro}
Carbyne is an infinitely long $sp$-hybridized linear chain of carbon atoms.\cite{heimann1999carbyne, Casari2018MCa} It is the last missing major member in the large family of carbon allotropes. With a cross-section of only one atom, carbyne is the prototypical example of a truly one-dimensional material.\cite{heimann1999carbyne} As such, it is expected to possess extraordinary properties, with its stiffness, strength, and thermal conductivity exceeding any known material, including other carbon-based materials like carbon nanotubes, graphene and diamond.\cite{Liu2013ANa, Casari2018MCa,Lechner2022CPBa} These prospects have motivated a large amount of theoretical studies on the properties of carbyne and experimental research towards its realization.\cite{candiotto2024strain, romanin2021dominant,  Wanko2016PRBa, Akagi1987SMa, Tykwinski2010PaACa, Kudryavtsev1993RCBa, Casari2018MCa, Cui2022AFMa, Feng2024NRa} Despite the strong and persistent scientific interest, theoretical studies, with e.g. density functional theory (DFT) calculations, remain challenging, which results in widely varying predictions for the properties of carbyne,\cite{Abdurahman2002PRBa,Yang2006TJoPCAa, Milani2007JoRSa, Mostaani2016PCCPa,Ramberger2021PCCPa,Shi2017PRMa,romanin2021dominant} e.g. for the band gap, where a range of values between 0.2 and 8 eV has been calculated.\cite{Yang2006TJoPCAa,Mostaani2016PCCPa} One of the main reasons for this ambiguity is the difficulty of modeling large systems with conjugated $\pi$-bonds by DFT accurately, as pointed out by Johnson \textit{et al.}\cite{Johnson2013TJoCPa} When modeling the potential energy surface of the fundamental stretching vibration of carbyne, DFT studies can even yield both a double and a single well potential depending on the chosen boundary conditions.\cite{candiotto2024strain}. Experimental data to anchor theoretical calculations and to verify or refute the predicted properties of carbyne is not available to date, because even after decades of efforts bulk carbyne has not been successfully synthesized.

While the synthesis of bulk carbyne has remained elusive, structurally similar materials exist, among them confined carbyne. It consists of long ($>$100 atoms) linear chains of $sp$-hybridized carbon atoms inside multi- or single-walled carbon nanotubes. Following earlier studies on linear carbon chains inside carbon nanotubes,\cite{Zhao2003PRLa, Cupolillo2008JoRSa} Shi \textit{et al}. reported isolated chains of confined carbyne comprising of hundreds and thousands of carbon atoms inside double-walled carbon nanotubes (DWCNTs).~\cite{shi2016confined} As the Peierls theorem predicts, confined carbyne features alternating single and triple bonds. This structure is described by the bond length alternation (BLA), which is defined as the length difference between short and long bonds. Confined carbyne possesses a single Raman active phonon mode, the so-called C mode, a longitudinal optical phonon that corresponds to an oscillation of the BLA along the chain axis. Different BLA values hence result in different C mode frequencies.\cite{heeg2018carbon} This makes Raman spectroscopy the primary experimental method to characterize confined carbyne, aided by the fact that this material possesses the highest Raman cross section ever reported.\cite{tschannen2020raman} In general, the C mode appears in the range between 1760 \ramcmspace and 1870 \ramcm.\cite{shi2016confined,heeg2018carbon}  It follows that carbyne chains with different BLAs exist inside carbon nanotubes. The nanotube host prevents the use of alternative optical techniques to characterize confined carbyne, like absorption spectroscopy due to overlapping resonances and photoluminescence spectroscopy due to quenching.\cite{Heeg2018Ca}

A previous extended Raman study on single isolated carbyne chains correlated the properties of a particular carbyne chain with its host nanotube.\cite{heeg2018carbon} Confined carbyne chains with a particular C mode frequency (and BLA) always reside inside the same host nanotube chirality. This shows that the BLA and related properties of confined carbyne chains are length-independent, but vary among different host nanotube chiralities. In particular, the diameter of the inner host nanotube acts as a parametrization of the properties of the linear carbyne chain.\cite{heeg2018carbon} Current strategies for extracting the properties of ideal carbyne from confined carbyne data focus on understanding and eventually factoring out this parametrization by the host nanotube.\cite{shi2016confined, heeg2018carbon}

Carbon atomic wires are the second major material class that is structurally similar to carbyne.\cite{marabotti2024synchrotron,Casari2016Na} These short linear chains of $sp$-hybrized carbon atoms are terminated by hydrogen or various other end groups. Their properties such as the frequency of their BLA oscillation, here called ECC mode from effective conjugation coordinate theory, are heavily dependent on the chain length.\cite{marabotti2024synchrotron} Hence, for carbon atomic wires the chain length acts as a parametrization of their properties, similar to the role of the host nanotube in confined carbyne. In addition to Raman spectroscopy, UV-vis absorption spectroscopy may also be applied in their study.\cite{Marabotti2022NCa} Current strategies for predicting the properties of ideal carbyne from carbon atomic wires focus on understanding the effect of chain length and extrapolating beyond these length dependencies.\cite{Tykwinski2010PaACa, Agarwal2016TJoPCCa, Gao2020NCa, Arora2023ACSa} 

While both short carbon atomic wires and confined carbyne possess similar molecular structure, to our knowledge there has been no attempt to combine their analysis to extract commonalities and unifying principles. This is likely due to the fact that both materials differ in a number of ways like synthesis methods, stability, and that they are mostly studied in disjunct scientific communities. On top, there is no obvious connecting property beyond a general agreement that both materials are related to ideal carbyne.\cite{shi2016confined, Tykwinski2010PaACa, Marabotti2022NCa} One pronounced experimental feature that appears in the Raman spectra of both materials are unusually strong vibrational overtones of the C and ECC mode, respectively.\cite{Marabotti2022NCa, Cupolillo2008JoRSa} Vibrational excitations are usually treated in the harmonic approximation, which yields equidistant quantized energy levels and overtones with frequencies that are exact multiples of the fundamental mode. Real vibrational potentials, however, deviate from the harmonic model. They give rise to anharmonic shifts of the energy levels and overtone frequencies, which often progressively increase for higher-order overtones.\cite{Morse1929PRa} Since anharmonicity is directly tied to the shape of the vibrational potential, it manifests a fundamental, intrinsic material property.

In this work, we establish a universal relationship between geometrical structure and vibrational anharmonicity in one-dimensional $sp$-hybridized carbon, further locking down the properties of ideal carbyne. To this end, we perform Raman spectroscopy of spatially separated confined carbyne chains up to the fourth order of the C mode (i.e., the 4C mode). In contrast to a previous study limited to the second vibrational order,\cite{Cupolillo2008JoRSa} we reveal a large degree of vibrational anharmonicity. Comparing these results to a similar work by Marabotti \textit{et al.} on short carbon atomic wires,\cite{marabotti2024synchrotron} we show that this vibrational anharmonicity is a unifying property of both confined carbyne and short carbon atomic wires by correlating anharmonicity with the material structure. As described previously, confined carbyne and carbon atomic wires are material systems with very different mechanisms of modification - while the BLA strongly varies with the chain length in carbon atomic chains,\cite{marabotti2024synchrotron} it is solely driven by the interaction with the surrounding nanotube host in confined carbyne and does not depend on the chain length.\cite{heeg2018carbon}
Therefore, we can exclude a direct influence of these factors and conclude that vibrational anharmonicity is intrinsically controlled by the chain geometry, characterized by the BLA, itself. Thus, by combining analysis of confined carbyne and short carbon atomic wires it is possible to
approach and approximate the properties of ideal carbyne.

\newpage
\section{Results}

\begin{figure}[h]
    \centering
    \includegraphics[width=8.9cm]{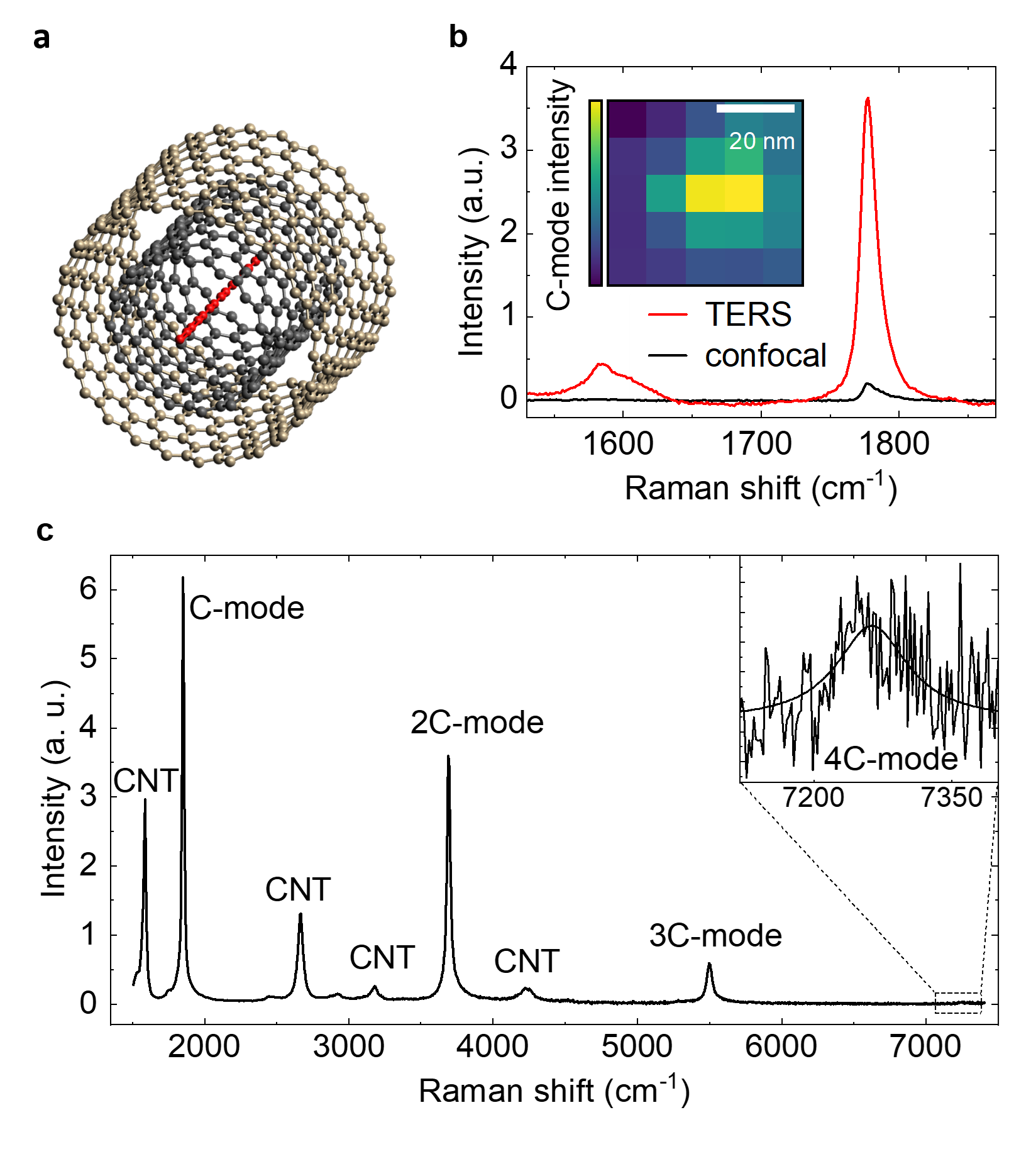}
    \caption{\textbf{TERS characterization and confocal Raman spectrum of an isolated confined carbyne chain with three C mode overtones} (a) Schematic representation of a confined carbyne chain inside a double-walled carbon nanotube (b) Tip-enhanced Raman spectroscopy (TERS) spectrum (red) compared to a confocal Raman spectrum (black) of an isolated confined carbyne chain. Both spectra are taken with 633 nm excitation. Inset: TERS map of C mode, revealing a chain length of $\sim$20 nm. (c) Confocal Raman spectrum of an isolated confined carbyne chain up to the fourth order of the C mode (4C). The 4C mode is magnified 20x due to its low intensity and a Lorentzian fit is included for clarity. The spectrum was taken using a 532 nm laser for excitation.}
    \label{fig:1}
\end{figure}
We measure the Raman spectra of isolated carbyne chains confined inside DWCNTs as sketched in Fig.~\ref{fig:1}a. The DWCNTs are dispersed on glass, allowing us to use tip-enhanced Raman spectroscopy to map the chain's C mode, Fig.~\ref{fig:1}b, and to confirm that the confined carbyne chains are indeed spatially separated. We focus on the Raman spectra of 16 confined carbyne chains with C mode frequencies ($\omega_{C}$) between 1786 - 1861 \ramcm. This covers around 70\% of the entire range of C mode frequencies reported for confined carbyne as a result of parametrization by the host nanotube, and ensures that we pick up any trend that depends on $\omega_{C}$.~\cite{Lechner2022CPBa,Heeg2018Ca,tschannen2020raman} Of these 16 chains, we detect three overtones of the fundamental C mode (i.e., up to the 4C mode) for 11 chains and two overtones (i.e., up to the 3C mode) for the 5 remaining chains. In order to accurately determine the frequencies of the C mode and its overtones, we only consider Raman spectra showing confined carbyne chains with well-resolved and thus clearly distinguishable peaks and exclude locations where the Raman spectrum shows major contributions of more than two carbyne chains. The corresponding C mode and overtone frequencies are provided in Section~S.4 and selection criteria in Section~S.1 of the Supporting Information. All Raman spectra are recorded with 532\,nm excitation, for which all confined carbyne chains studied here are excited at or close to their fundamental or higher vibronic resonances.~\cite{martinati2022electronic, Shi2017PRMa}

We show a representative Raman spectrum of an isolated confined carbyne chain in Fig.~\ref{fig:1}c. The chain's C mode appears as a singular peak at 1861 \ramcm. It is the only optically active phonon mode of confined carbyne, unlike in short carbon atomic wires, where additional termination-related modes appears.\cite{Tabata2006Ca} The first, second, and third overtones of the C mode, labelled 2C, 3C, and 4C, appear as approximate frequency multiples of the base C mode frequency. They originate from transitions to the higher energy levels of the chain's vibrational potential. Both the C mode and its overtones are of almost perfect Lorentzian shape such that they can be fitted with an accuracy better than 0.2\% of their frequency for all chains and modes reported here. The 2C mode appears at 3702 \ramcm, the 3C mode at 5505 \ramcm, and the 4C mode at 7251 \ramcm. As expected, the Raman intensity decreases with increasing overtone order. While the intensity of the 4$^{th}$ order is rather low compared to the noise level, the peak position can still be determined unambiguously. The clear separation and Lorentzian shape of the identified confined carbyne modes lets us conclude that there are no additional spectral features close to these peaks.

\begin{figure}[t]
    \centering
    \includegraphics[width=8.9cm]{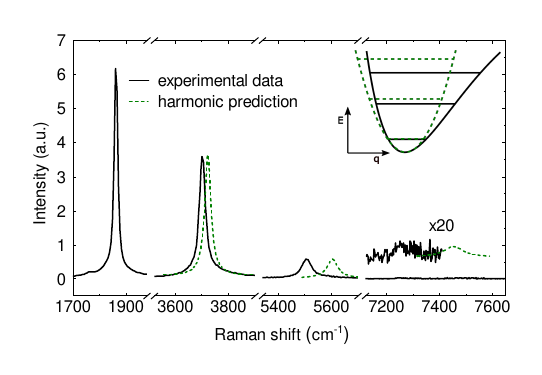}
    \caption{\textbf{Experimental anharmonic vs. hypothetical harmonic Raman spectrum of C mode and overtones.} Hypothetical spectrum based on a harmonic vibrational potential, compared with an experimental Raman spectrum of a confined carbyne chain with $\omega_{C}=1861$ \ramcm. The inset shows schematic representations of an anharmonic and a harmonic potential. The potential energy E is given in relation to a generic vibrational coordinate q.}
    \label{fig:2}
\end{figure}

We demonstrate the vibrational anharmonicity of confined carbyne  by comparing in Fig.~\ref{fig:2} the C mode and its overtones, extracted from Fig.~\ref{fig:1}c, to a hypothetical entirely harmonic Raman spectrum with equidistant peak spacing between two consequent vibrational energy levels. The increasing separation of the Raman peaks with frequency in the two spectra corresponds to a redshift of the experimentally observed overtones. This  shows the growing impact of the vibrational anharmonicity with higher-order C mode overtones. 

\begin{figure}[t]
    \centering
    \includegraphics[width=8.9cm]{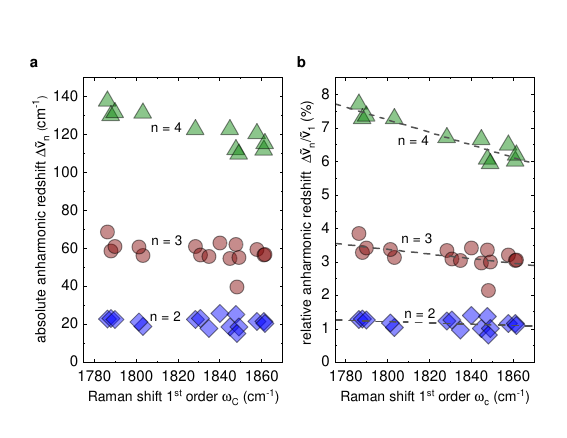}
    \caption{\textbf{Anharmonic redshift of the C mode overtones of 16 confined carbyne chains.} The 2C and 3C mode could be recorded for all chains, the 4C mode for 11 chains. (a) Absolute anharmonic redshift $\Delta\widetilde{\nu}_{n} = \widetilde{\nu}_{1} - (\widetilde{\nu}_{n} - \widetilde{\nu}_{n-1})$ and (b) relative anharmonic redshift $\Delta\widetilde{\nu}_{rel,n} = \Delta\widetilde{\nu}_{n}/\widetilde{\nu}_{1}$ as a function of the fundamental C mode frequency $\omega_{C}$.}
    \label{fig:3}
\end{figure}

The spacing between two consecutive vibrational energy levels is given by $\widetilde{\nu}_{n}-\widetilde{\nu}_{n-1}$, where $\widetilde{\nu}_{n}$ is the vibrational frequency of the n$^{th}$ carbyne phonon mode nC. In a harmonic vibrational potential all spacings between two neighboring energy levels have the same magnitude, yielding the relation $\widetilde{\nu}_{n}-\widetilde{\nu}_{n-1} = \widetilde{\nu}_{1}$ for all n. In contrast, in an anharmonic potential with redshifted overtones, we obtain $ \widetilde{\nu}_{n}-\widetilde{\nu}_{n-1} < \widetilde{\nu}_{1} $. Thus, to quantify the anharmonicity in confined carbyne, we define the anharmonic redshift $\Delta\widetilde{\nu}_{n}$ as $\Delta\widetilde{\nu}_{n} = \widetilde{\nu}_{1} - (\widetilde{\nu}_{n} - \widetilde{\nu}_{n-1})$. We extract $\Delta\widetilde{\nu}_{n}$ from our experimental data and plot it as a function of C mode frequency in Fig.~\ref{fig:3}a for 11 confined carbyne chains up to the 4C mode and for 5 confined carbyne chains up to the 3C mode. Fig.~\ref{fig:3}a demonstrates the absolute magnitude of vibrational anharmonicity. Ignoring any dependence of $\Delta\widetilde{\nu}_{n}$ on the C mode frequency at this point, we find the expected  general increase in relative anharmonicity with overtone order. However, to compare the anharmonicity of chains with different C mode frequencies, we renormalize the absolute anharmonic shift by the base C mode frequency and define the relative anharmonic redshift $\Delta\widetilde{\nu}_{rel,n} = \Delta\widetilde{\nu}_{n}/\widetilde{\nu}_{1}$. $\Delta\widetilde{\nu}_{rel,n}$ only depends on the deviation of the shape of the vibrational potential from the harmonic approximation. 

To quantify the vibrational anharmonicity in confined carbyne, we plot the relative anharmonic redshift $\Delta\widetilde{\nu}_{rel,n}$ as a function of C mode frequency and overtone order for all 16 chains in Fig.~\ref{fig:3}b and make two key observations. 
First, the relative anharmonic redshift increases strongly with the mode order, reaching 6.0-7.7$\%$ for the 4C mode. These values are up to 8 times larger compared to other solid state systems for which a similar number of overtones is reported (GaN: 0.95$\%$, ZnO: 1.20$\%$, ZnTe: 1.67$\%$)~\cite{Sun2002JoAPa,Callender1973PRBa,Zhang1993PRBa}. Second, we find a clear trend that connects the C mode frequency and the anharmonic redshift. Confined carbyne chains with lower C mode frequency possess greater vibrational anharmonicity. The trend is indicated by the dashed lines in Fig.~\ref{fig:3}b. While the trend is masked for the 2C mode by what appears to be observational random error, it can be made out in the 3C mode and is clearly pronounced in the 4C anharmonic redshift. The anharmonic redshift increases from around 6.0$\%$ for the chain with $\omega_{C}$ = 1861 \ramcm$ $  to 7.7$\%$ for the chain with  $\omega_{C}$ = 1786 \ramcm, which represents a relative change of 28$\%$.

Since the C mode frequency of confined carbyne is directly tied to its bond length alteration and smaller frequencies correspond to a smaller BLA, we conclude that the vibrational anharmonicity in confined carbyne increases with decreasing C mode frequency/BLA. The BLA directly determines a carbyne chain's cumulenic or polyynic character, with a BLA of zero corresponding to a perfect cumulenic state. This means that the more evenly the $\pi$-electrons are distributed along the chain, the higher the vibrational anharmonicity of confined carbyne is.

\section{Discussion}
To gain further insight into the strongly anharmonic behavior observed in confined carbyne's C mode overtones, we model the anharmonic redshift and extract the corresponding anharmonic nondimensional parameter ($\chi$) using second-order vibrational perturbation theory (VPT2) as formulated by Mendolicchio \textit{et al.} \cite{mendolicchio2022perturb} VTP2 successfully captures the anharmonic behavior of short carbon atomic wires (\ie hydrogen-capped polyynes)~\cite{marabotti2024synchrotron}. Given the structural and vibrational similarities between confined carbyne and short carbon atomic wires, we apply the same model to estimate their vibrational anharmonicity. Importantly, VTP2 does not rely on a particular potential because it describes the degree of vibrational anharmonicity without defining a specific potential shape. This is particularly beneficial for carbyne, since the exact shape of the potential energy surface of the BLA oscillation in carbyne is not known.  Carbyne has been modeled both with single and double minima potential wells by different authors.~\cite{candiotto2024strain, artyukhov2014mechanically, Milani2007JoRSa} The corresponding DFT calculations are not conclusive since both models predict unrealistic C mode frequencies far off the experimental data.~\cite{candiotto2024strain, artyukhov2014mechanically, milani2008first} Therefore, our discussion focuses on describing and quantifying the observed anharmonic trend in confined carbyne such that our conclusions remain valid for any theoretical model or potential shape of choice to describe carbyne. 

\begin{figure}[t]
    \centering
    \includegraphics[width=16.5cm]{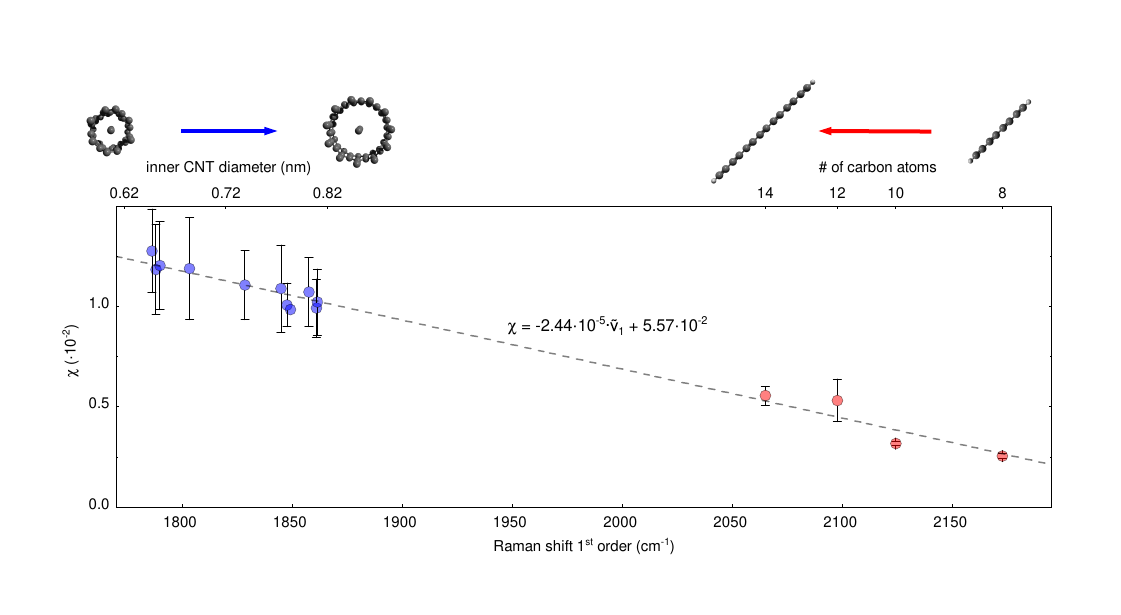}
    \caption{\textbf{Universal vibrational anharmonicity  of linear carbon} Anharmonicity ($\chi$) as a linear function of the first-order BLA oscillation Raman mode frequency for confined carbyne (C mode, blue dots) and carbon atomic wires (ECC mode, red dots). The errors derive from the VTP2 fit. The linear correlation (dashed black line) between $\chi$ and the Raman mode frequency follows Eq.~\ref{eq:universal_chi}. The values of the inner nanotube diameter, which parametrizes the  C mode/BLA of confined carbyne chains, are calculated from Eq.~1 in Ref.\citer{heeg2018carbon}. The correlation between the chain lengths of carbon atomic wires  and the ECC mode/BLA is described in Ref.\citer{marabotti2024synchrotron}.}
    \label{fig:chi}
\end{figure}

Since confined carbyne features only one Raman-active mode, we simplify the expression of the vibrational energy levels ($E_n$) of Ref.\citer{mendolicchio2022perturb} and obtain

\begin{equation}
    \frac{E_n}{hc} = \varepsilon_0 + \Tilde{\nu}_{harm} n - \Tilde{\nu}_{harm} \chi \left( n^2 + n \right),
    \label{eq:vibr_en_lev_VPT2}
\end{equation}

where $\varepsilon_0$ is the vibrational zero-point frequency (in~\ramcm), $\Tilde{\nu}_{harm}$ is the C mode frequency within the harmonic approximation of the potential energy surface (in~\ramcm), and $n$ is the order of the vibration level. Here $\Tilde{\nu}_{harm}$ represents a renormalization constant to express $\chi$ in nondimensional units. Further details on the  derivation of Eqn.~\ref{eq:vibr_en_lev_VPT2} is given in Section~S.2 of the Supporting Information.

We extract the values of $\chi$ from a linear regression of the experimental frequency spacings between subsequent vibrational levels for the 11 confined carbyne chains that feature three overtones, see Table S.1 for tabulated values and Eq.~S.3 in the Supporting Information. Figure~\ref{fig:chi} shows the resulting $\chi$ values as a function of C mode frequency for confined carbyne (blue points). Before discussing and interpreting our data in detail, it must be noted that the $\chi$ values for confined carbyne shown in Fig.~\ref{fig:chi} have significant error bars. This indicates that VPT2 provides a quick and direct estimate of vibrational anharmonicity but does not fully account for the stark increase in the anharmonic redshift with overtone order. Fourth-order vibrational perturbation theory (VPT4) as established by Gong \textit{et al.}\cite{gong2018fourth} yields a better fit to our experimental data than VTP2, as discussed in detail in Section~S.3 of the Supporting Information. The complex formulation of VPT4's two anharmonic parameters, however, complicates the interpretation while keeping the conclusions unchanged compared to VPT2. Therefore, we focus on the VPT2 results but provide our experimental data for fitting  with VPT4 or other theoretical models that describe vibrational anharmonicity.

Figure~\ref{fig:chi} clearly shows that the vibrational anharmonicity of confined carbyne increases as the C mode frequency decreases. This means that the anharmonicity is driven by the BLA, as $\chi$ increases as the BLA (or BLA oscillation frequency) reduces. Previously, we found a roughly linear relationship between the diameter of the innermost host nanotube and the C mode frequency of the confined carbyne chain, where a larger (smaller) inner nanotube diameter corresponds to a higher (lower) C mode frequency~\cite{Heeg2018Ca,heeg2018carbon}. Combining this relationship with our results shows that the host nanotubes act as a parametrization of both the chain's BLA and the corresponding vibrational anharmonicity. The host nanotube diameters are provided in Fig.~\ref{fig:chi} for the relevant C mode range.

Next, we extract the vibrational anharmonicity  of short carbon atomic wires from the Raman data  reported in Ref.\citer{marabotti2024synchrotron}. The $\chi$ values for carbon atomic wires (red points) are shown in Fig.~\ref{fig:chi} as a function of their ECC mode~\cite{marabotti2024synchrotron}, which denotes the BLA oscillation in H-capped polyynes. The ECC mode frequency redshifts with increasing chain length, corresponding to a reduction of BLA. As for confined carbyne, the vibrational anharmonicity of carbon atomic wires increases with decreasing BLA/ECC mode frequency, albeit at lower total values than in confined carbyne. 

Upon comparing how $\chi$ depends on the C mode for confined carbyne and the ECC mode for carbon atomic wires, we find that the vibrational anharmonicity of the BLA oscillation follows a universal law applicable to both systems. The anharmonicity $\chi$ changes as a linear function of the BLA oscillation frequency $\widetilde{\nu}_1$, sketched as a dashed line in Fig.~\ref{fig:chi}, and follows 

\begin{equation}
    \chi (\widetilde{\nu}_1) = a \cdot \widetilde{\nu}_1 + b,
    \label{eq:universal_chi}
\end{equation}

where $ a = -2.44 \pm 0.09 \left[ \cdot 10^{-5} \frac{1}{cm^{-1}} \right] $ and $ b = 5.57 \pm 0.18 \left[ \cdot 10^{-2} \right] $. This universal dependence of the anharmonicity on the BLA is the first property directly linking confined carbyne chains, governed by the Peierls distortion, and isolated short carbon atomic wires, whose properties are determined by size confinement effects. This shows that a specific environment such as length confinement or nanotube encapsulation affects the BLA and BLA oscillation frequency of an $sp$-hybridized structure, but does not alter the structure's intrinsic physical properties and their scaling. 

Our observations further suggest that any other carbyne-like material systems will show the same dependence of the vibrational anharmonicity on the BLA.  The corresponding $\chi$ values can be predicted using Eq.~\ref{eq:universal_chi} based on the experimentally observed BLA oscillation frequency $\widetilde{\nu}_1$. This has several important implications, which we will discuss in the following. 
 
First, we argue that the universal relationship between the vibrational anharmonicity and BLA oscillation frequency is a good measure for experimentally characterizing carbyne-like material systems. If a material shows the vibrational anharmonicity predicted by Eq.~\ref{eq:universal_chi} in its overtone Raman spectra, it is a  strong indication that the material is indeed a carbyne-like material. Beyond anharmonicity, the BLA oscillation frequency $\widetilde{\nu}_1$ emerges as a reliable and easily measurable parameter to reveal intrinsic relationships between the properties of a carbyne-like material and its BLA, without the need for direct measurement of the BLA, which requires high-purity crystalline samples. By establishing this relationship, we can infer information about the BLA from easily obtainable Raman measurements, simplifying the study of these materials, irrespective of their specific environment. For instance, if we can relate a specific property of either confined carbyne chains or short carbon atomic wires to their BLA oscillation frequency, we can, in principle, predict the behavior of all carbyne-like systems.
 
Second, we can infer the vibrational anharmonicity of ideal carbyne. Since ideal carbyne can be understood both as confined carbyne of infinite diameter host nanotube and as a carbon atomic wire with infinite length, the corresponding $\widetilde{\nu}_1$ and $\chi$ values represent upper and lower limits, respectively. A recent experimental work on carbon wires up to 52 atoms long and stabilized by complexation with platinum-based endgroups extrapolated $\widetilde{\nu}_1 = 1881$ cm$^{-1}$ for ideal carbyne,\cite{Arora2023ACSa} which lies within the expected range. Based on this value, Eq. \ref{eq:universal_chi} predicts a vibrational anharmonicity of $\chi = 0.98\cdot 10^{-2}$ for ideal carbyne.

Third, we believe that the universal relationship between BLA oscillation frequency and vibrational anharmonicity reported here is of considerable value to theoreticians, providing a basis for refined models and achieving more accurate predictions. The $\chi$ values provided here may serve as a benchmark for theoretical models describing ideal carbyne as well as confined carbyne, carbon atomic wires and other finite realizations of carbyne. Note that to date, only few theoretical studies have accounted for anharmonic effects in carbyne,\cite{romanin2021dominant} while most neglected them, assuming they are small.\cite{Abdurahman2002PRBa,Yang2006TJoPCAa,Mostaani2016PCCPa,Ramberger2021PCCPa} The comparably large vibrational anharmonicity reported here calls this approach into question. In agreement with large electron-phonon interactions observed by Martinati \textit{et al.}\cite{martinati2022electronic} for confined carbyne, our findings suggest that carbyne may actually possesses a lower thermal conductivity and electron mobility than previously predicted.~\cite{ferreira2020direct, Wang2015SRa, iqbal2019enhanced, qin2020multiphonon}. Future studies will have to consider anharmonic effects and reevaluate how this affects other properties of carbyne, such as its record mechanical strength.\cite{Gao2020ANa}

In conclusion, we have quantified the vibrational anharmonicity of confined carbyne by measuring the Raman modes of single confined carbyne chains up to the third overtone. We find a vibrational anharmonic redshift that increases with overtone order, reaching $8\%$ for the third overtone, and a monotonic increase in vibrational anharmonicity with decreasing C mode frequency. Comparing different confined carbyne chains with short carbon atomic wires of varying length, we uncover a universal linear relation between the BLA oscillation frequency and the vibrational anharmonicity of carbyne-like, linear one-dimensional carbon systems. This relationship serves as a fingerprint to identify carbyne-like materials by Raman spectroscopy and acts as an experimental anchor and  benchmark for refining theoretical models describing carbyne and related materials. Our work shows that the vibrational anharmonicity of carbyne-like systems cannot be neglected in their theoretical description. 
\section{Methods}
Carbyne chains were grown inside DWCNTs in a high-temperature, high-vacuum process as described in Ref.\citer{shi2016confined} and dispersed on a thin glass coverslip. Confocal Raman spectroscopy measurements are conducted with an Xplora Raman spectrometer (Horiba) using 532 nm excitation. We use acquisition times between 10~s and 150~s, depending on the intensity of the Raman response of specific chains. The applied excitation laser power was kept below 5 mW. Following Ref.\citer{Tschannen2021ANa}, we expect heating of at most 25 K, which corresponds to a heating induced shift of the C mode smaller than 0.6 \ramcm. Measuring two chains with very similar C mode frequency (difference $<0.4$ \ramcm) at different powers (1.4 and 4.9 mW respectively), we obtain values of $\chi$ within 3$\%$ of each other. This leads us to conclude that heating effects can be neglected.
We use a neon glow lamp to calibrate the spectral position data. TERS measurements were conducted with a FabNS Porto-SNOM using a 633 nm He-Ne excitation laser and plasmon-tunable tip pyramids~\cite{Vasconcelos2018AOMa}.

\section{Acknowledgements}
J.M.A.~L., P.~M. and S.~H. acknowledge funding from the Deutsche Forschungsgemeinschaft (DFG) under the Emmy
Noether Initiative (Project-ID 433878606). P.~M. acknowledges the financial support of the Einstein International Postdoctoral Fellows program (IPF-2022-727). L.~S. acknowledges the National Natural Science Foundation of China (52472059). 

\section{Author Contributions}
J.M.A.~L. performed the confocal Raman measurements with support from P.~M. and evaluated the experimental data. P.~M. modelled the vibrational anharmonicity and performed the TERS measurements. L.~S. and T.~P. provided the confined carbyne sample. C.S.~C. assisted in data interpretation and modelling. J.M.A.~L., P.~M., and S.~H. interpreted the data and co-wrote the manuscript with input from all authors. S.H. conceived and supervised the project.

\bibliography{references}

\end{document}


\tableofcontents
\section{Raman spectra of confined carbyne chains and selection criteria}
\label{sec:ram_spec}
\begin{figure}[!b]
    \centering
    \includegraphics[width=\linewidth]{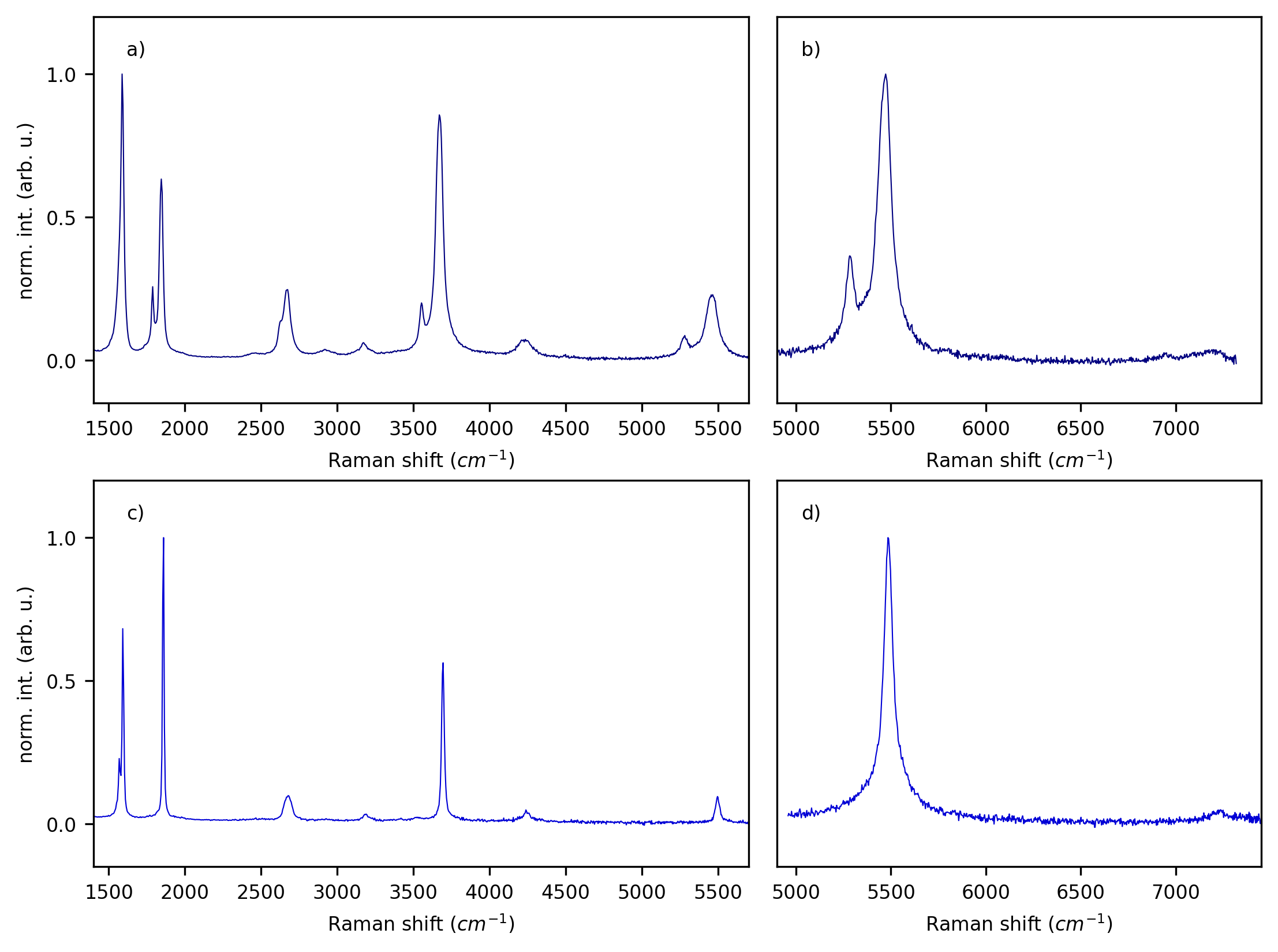}
    \caption{ \textbf{Raman spectra of confined carbyne chains at two separate locations.} (a) and (b) show spectra of a location where two confined carbyne chains are probed within the laser spot. Their C-mode and overtones can  be spectrally resolved unambiguously. (c) and (d) show spectra of a location with a single isolated confined carbyne chain. (a) and (c) show the C, 2C and 3C modes, while (b) and (d) show the 3C and 4C modes of the corresponding chain(s). A separately measured background has been substracted from all spectra.}
    \label{fig:SI1}
\end{figure}
Typical Raman spectra of the confined carbyne chains measured for this work are presented in Fig. \ref{fig:SI1}. Since the spectrometer used does not capture the whole spectral bandwith from the C mode to the 4C mode in a single spectrum, two spectra are recorded for each location where the 4C mode could be detected. The first spectrum, \ref{fig:SI1}(a) and (c), captures the C, 2C and 3C modes, while the second spectrum, \ref{fig:SI1}(b) and (d), captures the 3C and 4C mode.

Other than the C mode and its overtones, several more peaks appear in the measured spectral range, which originate from the host nanotubes. The well known G mode of CNTs can be found just below 1600 \ramcm, while the 2D mode appears at about ~2700 \ramcm.\cite{reich2008carbon} At roughly 3200 \ramcmspace the second order of the G mode also appears. The G + 2D combination mode is visible at ~4300 \ramcm. In some spectra a small peak at ~2900 \ramcmspace appears, which has been assigned to either a D + D' combination mode or a double-resonant overtone of a $q \neq 0$ LO phonon.\cite{Fantini2008TJoPCCa,Dresselhaus2005PRa}

\begin{figure}[!t]
    \centering
    \includegraphics[width=8.9cm]{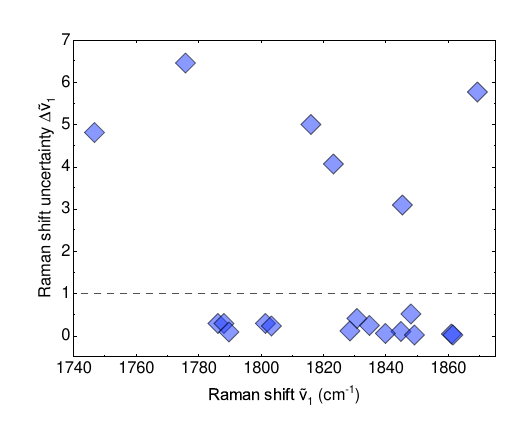}
    \caption{\textbf{Positional fitting uncertainty $\Delta\widetilde{\nu}_{1}$} of the Lorentzian functions used to fit the C-mode of confined carbyne Raman spectra. If the positional uncertainty $\Delta\widetilde{\nu}_{1} > 1$ \ramcm, the corresponding peak is excluded from our analysis.}
    \label{fig:SI2}
\end{figure}

We determine the Raman shifts of the confined carbyne Raman peaks with Lorentzian fit functions. In order to quantify the anharmonicity in confined carbyne with high precision, only chains with well-resolved peaks with negligible fitting error are included in our analysis. We consider a confined carbyne chain well-resolved if the fitting uncertainty corresponding to the spectral position of the Lorentzian fit function of its C mode is smaller than 1 \ramcm.  This parameter is plotted in Fig. \ref{fig:SI2} for the functions used to fit the C mode of all chains. This value is somewhat arbitrary but it is comparable to the resolution of scientific grade Raman spectrometers. Further, this value separates a group of chains with C modes that gave very little fitting uncertatinty from  chains whose C modes can only be fitted well with larger errors of several \ramcmspace.

Additionally, spectra that show major contributions from more than 2 chains are excluded to decrease fitting uncertainty and prevent erroneous attribution of overtones to the wrong confined carbyne chain. This is considered to be the case if the integrated area of the third-largest Lorentzian fit function used for a certain peak is at least 10$\%$ as large as the integrated area of the second-largest Lorentzian fit function.

Based on the selection criteria described above, the Raman shift of the C mode and overtones of the confined carbyne chains used for this study are reported in Table \ref{tab:all} together with the corresponding values of carbon atomic wires from Ref.\citer{marabotti2024synchrotron}.

\section{Evaluation of anharmonicity of BLA oscillation using VPT2} \label{sec:vpt2_si}

In the framework of second-order vibrational perturbation theory (VTP2)~\cite{mendolicchio2022perturb}, anharmonic vibrational levels can be derived by introducing an anharmonic correction matrix ($\mathbf{\chi}$). This method describes the perturbed potential energy curve as a fourth-degree polynomial. In the single-mode approximation and considering no mode mixing (as confirmed by our experimental data, see the Results section in the main text), valid for the ideal carbyne system and confined carbyne, the harmonic expression of the potential energy is corrected by adding terms proportional to $q^3$ and $q^4$, where $q$ is the vibrational normal coordinate. The general expression for the vibrational level of quantum number $n$ is expressed by Eq.~1 in the main text and reported here for convenience:

\begin{equation}
    \frac{E_n}{hc} = \varepsilon_0 + \Tilde{\nu}_{harm} n - \Tilde{\nu}_{harm} \chi \left( n^2 + n \right),
\end{equation}

where $\varepsilon_0$ is the zero-point vibrational energy, $\Tilde{\nu}_{harm}$ is the harmonic frequency (in \ramcm) of the vibrational mode, and $\chi$ is the nondimensional anharmonic  parameter. For the difference between the $(n+1)^{th}$ and $n^{th}$ levels, we obtain

\begin{equation}
    \frac{E_{n+1} - E_n}{hc} = \Tilde{\nu}_{harm} \left( 1 - 2\chi \right) - \left( 2\chi\Tilde{\nu}_{harm} \right) n = K-Sn.
    \label{eq:linear_model}
\end{equation}
\begin{figure}[!t]
    \centering
    \includegraphics[width=\linewidth]{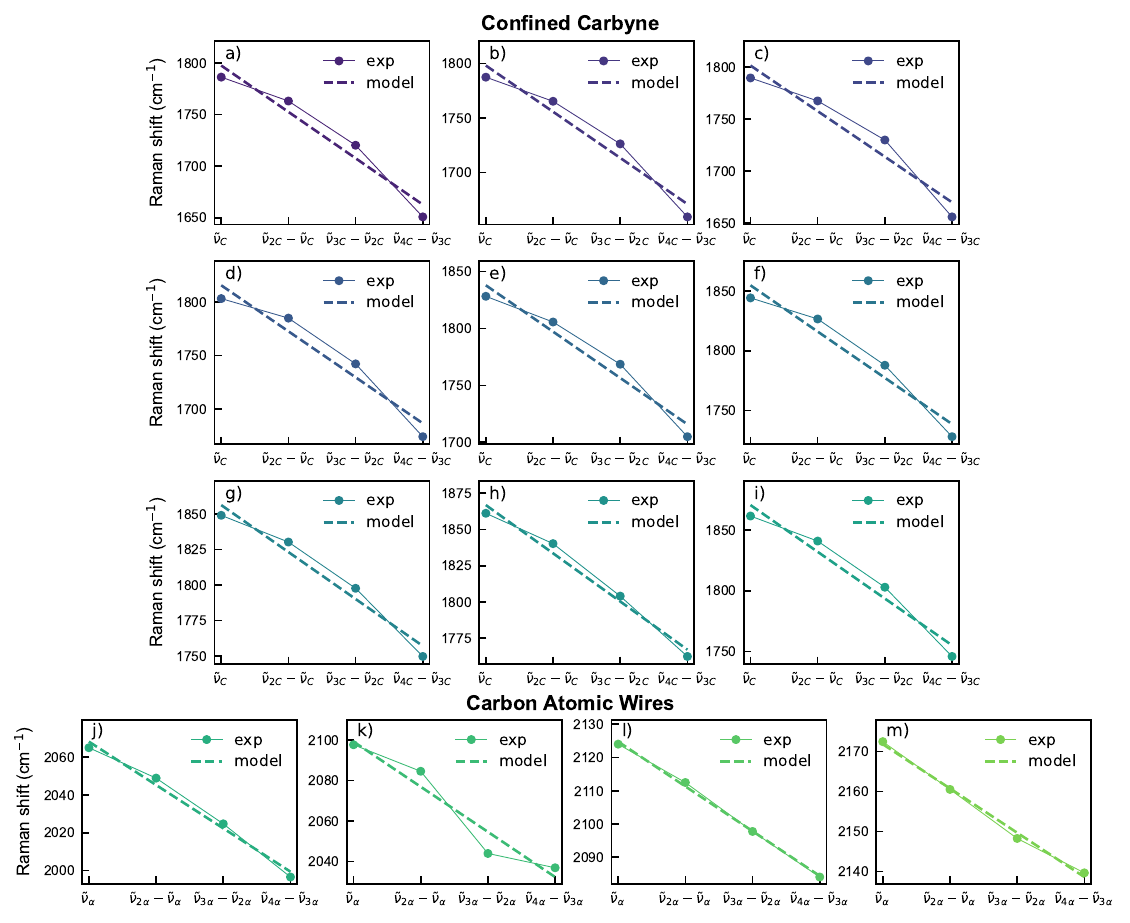}
    \caption{\textbf{Modeling of anharmonic redshifts according to VPT2} a-i) Frequency spacing between two subsequent vibrational quanta of the C mode (colored circles, ($\Tilde{\nu}_{(n+1)C} - \Tilde{\nu}_{nC}$, with n=1,2,3,...) and results of the linear fit procedure (dotted lines, see Eq.~\ref{eq:linear_model}) for confined carbyne. j-m) Frequency spacing between two subsequent vibrational quanta of the $\alpha$ mode (colored circles, ($\Tilde{\nu}_{(n+1)\alpha} - \Tilde{\nu}_{n\alpha}$, with n=1,2,3,...) and results of the linear fit procedure (dotted lines, see Eq.~\ref{eq:linear_model}) for carbon atomic wires from Ref.\citer{marabotti2024synchrotron}.}
    \label{fig:VPT2_fit}
\end{figure}
This relation relies exclusively on parameters that can be determined experimentally by Raman spectroscopy and does not require assuming values that cannot be determined, such as, i.e., the zero-point vibrational energy. Furthermore, we can use Eq.~\ref{eq:linear_model} to model our experimental data with a linear regression. From the intercept ($K$) and the slope ($S$), we can calculate $\Tilde{\nu}_{harm}$ and $\chi$ as 

\begin{equation}
    \begin{aligned}
        & \Tilde{\nu}_{harm} = K + S  \\
        & \chi = \frac{1}{2 \left( 1 + K/S \right)}.
    \end{aligned}
    \label{eq:linear_chi}
\end{equation}

Among the 16 confined carbyne chains available, we selected only 11 for which we can detect up to the third overtone (\ie 4C mode). Their C mode frequencies are listed in Table~\ref{tab:all}. In this work, we compared confined carbyne chains to short carbon atomic wires, whose Raman spectra and the frequencies of their corresponding $\alpha$ mode and overtones are displayed in Ref.\citer{marabotti2024synchrotron}. Figure~\ref{fig:VPT2_fit} illustrates the experimental frequency spacings between subsequent vibrational levels in confined carbyne chains and short carbon atomics wires, along with the results of linear regression, while the values of $\Tilde{\nu}_{harm}$ and $\chi$ are reported in Table~\ref{tab:all}.

\section{Evaluation of anharmonicity of BLA oscillation using VPT4}

As can be evinced from Figure~\ref{fig:VPT2_fit}, a linear fit based on VPT2 cannot fully capture the super-linear decrease in vibrational energy spacing of confined carbyne chains. To increase the degree of the polynomial fit (\ie quadratic function), we have to introduce an additional anharmonic parameter. Considering vibrational perturbation theory, this implies passing from VPT2 to VPT4 where a sixth-degree polynomial ($q^5$ and $q^6$) has been used to model the perturbed potential energy curve~\cite{gong2018fourth}. To describe the general expression for the energy of the vibrational level of quantum number $n$, we will use the rationalization of VPT4 as established by Gong \etal~\cite{gong2018fourth}, where now two anharmonic correction matrices appear ($\mathbf{\chi}$ and $\mathbf{\psi}$). Using the same approximations employed for VPT2 in Section~\ref{sec:vpt2_si}, we obtain

\begin{equation}
    \frac{E_n}{hc} = \varepsilon_0 + \Tilde{\nu}_{harm} n + \chi \left( n^2 + n \right) + \psi \left( n^3 + \frac{3}{2} n^2 + \frac{3}{2} n \right).
\end{equation}

The different sign in front of the linear term depending on $\chi$ compared to VPT2 implies that in VPT4 the first-order anharmonic correction may bring either a positive or negative contribution to the harmonic frequency. Here, the difference between the $(n+1)^{th}$ and $n^{th}$ levels becomes

\begin{equation}
    \frac{E_{n+1} - E_n}{hc} = \Tilde{\nu}_{harm} \left(1 + 2\chi + \frac{13}{4}\psi \right) + \left( 2\chi + 6\psi \right) \Tilde{\nu}_{harm} n + \left( 2\psi \right) \Tilde{\nu}_{harm} n^2 = K + Sn + Rn^2.
    \label{eq:quadratic_model}
\end{equation}

Using a quadratic curve to model our experimental data, from the set of fitting parameters ($K$, $S$, and $R$), we can extract $\Tilde{\nu}_{harm}$, $\chi$, and $\psi$:

\begin{equation}
    \begin{aligned}
        & \Tilde{\nu}_{harm} = K - S + \frac{11}{12} R \\
        & \chi = \frac{12R - 6S}{11R - 12S + 12K} \\
        & \psi = \frac{4R}{11R - 12S + 12K}
    \end{aligned}
    \label{eq:quadratic_chi_psi}
\end{equation}

\begin{figure}[!t]
    \centering
    \includegraphics[width=\linewidth]{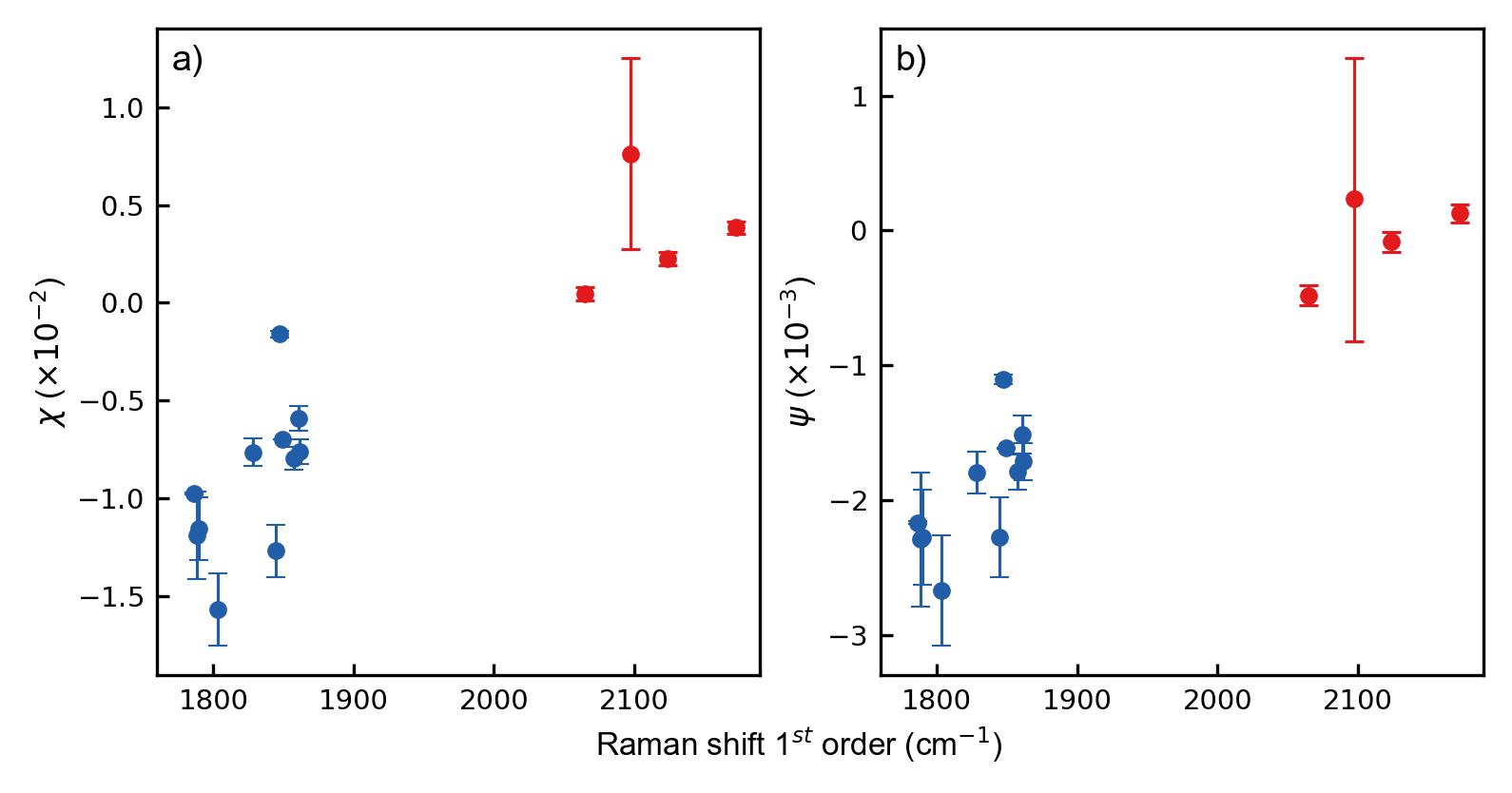}
    \caption{\textbf{a) First ($\chi$) and b) second-order ($\psi$) anharmonic parameters} of confined carbyne chains (blue points) and carbon atomic wires (red points) as a function of the Raman shift of their corresponding first-order mode according to VPT4. The errors derive from the fit (see Eqs.~\ref{eq:quadratic_model} and~\ref{eq:quadratic_chi_psi}).}
    \label{fig:VPT4_chi_psi}
\end{figure}
\begin{figure}[!t]
    \centering
    \includegraphics[width=\linewidth]{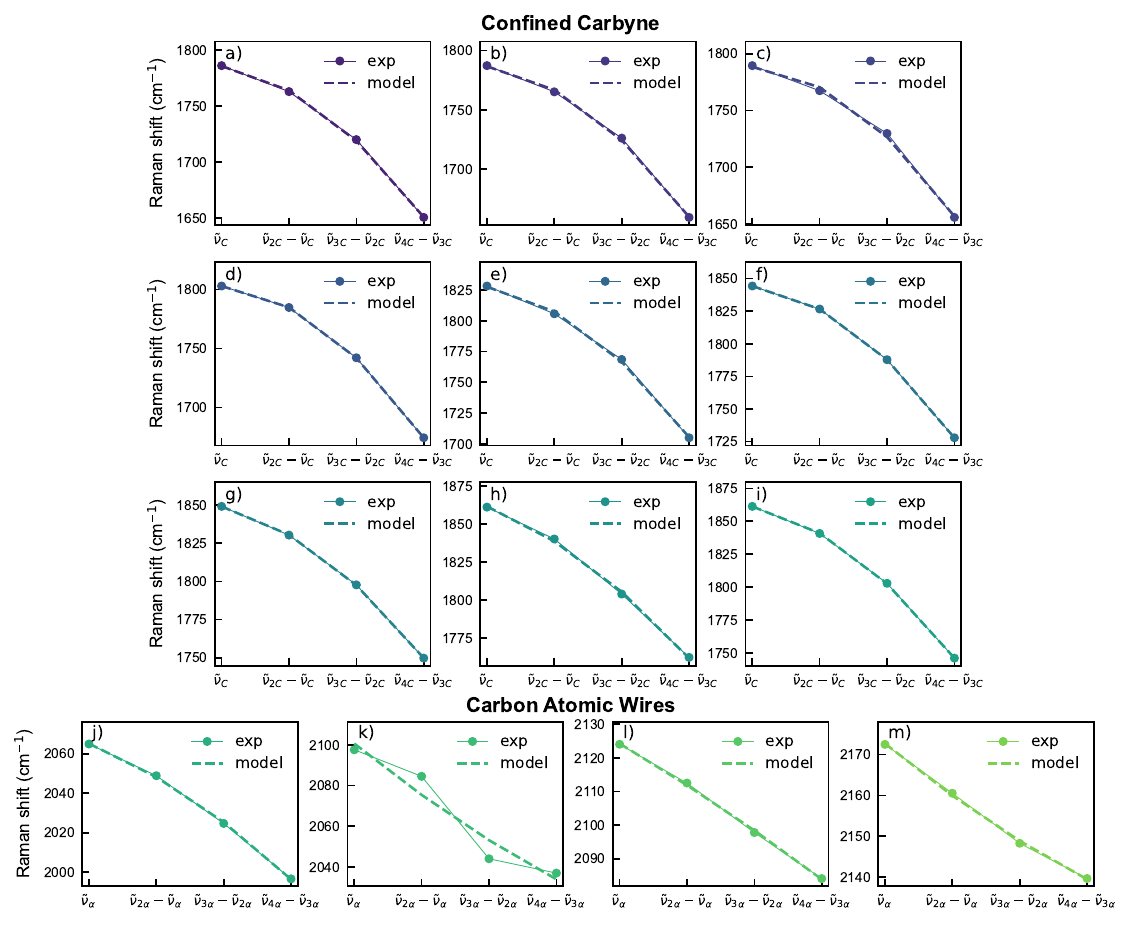}
    \caption{ \textbf{Modeling of anharmonic redshifts according to VPT4} a-i) Frequency spacing between two subsequent vibrational quanta of the C mode (colored circles, ($\Tilde{\nu}_{(n+1)C} - \Tilde{\nu}_{nC}$, with n=1,2,3,...) and results of the linear fit procedure (dotted lines, see Eq.~\ref{eq:quadratic_model}) for confined carbyne. j-m) Frequency spacing between two subsequent vibrational quanta of the $\alpha$ mode (colored circles, ($\Tilde{\nu}_{(n+1)\alpha} - \Tilde{\nu}_{n\alpha}$, with n=1,2,3,...) and results of the linear fit procedure (dotted lines, see Eq.~\ref{eq:quadratic_model}) for carbon atomic wires from Ref.\citer{marabotti2024synchrotron}.}
    \label{fig:VPT4_fit}
\end{figure}

Figure~\ref{fig:VPT4_fit} illustrates the experimental frequency spacings between subsequent vibrational levels in confined carbyne chains and short carbon atomics wires, along with the results of the regression, while the values of $\Tilde{\nu}_{harm}$, $\chi$, and $\psi$ are reported in Table~\ref{tab:all} and Figure~\ref{fig:VPT4_chi_psi}.

The results from modeling with VPT4 the vibrational energy spacings of confined carbyne and carbon atomic wires indicate a clear trend: both $\chi$ and $\psi$ parameters decrease with decreasing BLA oscillation frequency. Since the BLA oscillation frequency is linked to the BLA, our findings with VPT4 align with the VPT2 results, indicating that the anharmonicity of carbyne-like systems increases as their BLA decreases. The quadratic fitting of VPT4 aligns better with the experimental data, particularly for confined carbyne chains, as is evident when comparing Fig.~\ref{fig:VPT2_fit} and Fig.~\ref{fig:VPT4_fit}. Specifically, in confined carbyne, both $\chi$ and $\psi$ are negative, meaning that their vibrational energy spacings deviate from the harmonic scheme and they get smaller as the BLA increases (see Eq.~\ref{eq:quadratic_model}). In contrast, carbon atomic wires feature positive $\chi$ for all chain lengths analyzed here, and $\psi$ slightly negative, nearly zero for shorter chains. This distinction highlights the enhanced anharmonic behavior of confined carbyne compared to short carbon atomic wires, emphasizing their greater degree of anharmonicity. 

However, the interplay between these two anharmonic parameters in defining the anharmonic vibrational spacings is complex and not well understood. A positive $\chi$ indicates that the quadratic term with $n$ (\ie vibrational level) in Eq.~\ref{eq:quadratic_model} is positive, increasing the vibrational energy spacing. This contradicts the common observation that vibrational anharmonicity decreases the harmonic vibrational energy spacing. Conversely, the cubic term, modulated by $\psi$, brings a negative contribution. Nevertheless, drawing conclusions beyond this phenomenological analysis is challenging. Due to this complexity and the poorly theoretically explored interplay between the anharmonic parameters, we use VPT2 to address anharmonicity in carbyne-like systems. 

\newpage
\section{Experimental data and anharmonic parameters}


\begin{table}[]
    \centering
    \begin{tabular}{c|c|c|c|c|c|c|c|c}
        \hline \hline
        \multicolumn{9}{c}{\textbf{Confined Carbyne}} \\
        \hline
        \multicolumn{4}{c|}{Raman shift} & \multicolumn{2}{c|}{VPT2} & \multicolumn{3}{c}{VPT4} \\
        \hline
        
         $\Tilde{\nu}_{1C}$& $\Tilde{\nu}_{2C}$ & $\Tilde{\nu}_{3C}$ & $\Tilde{\nu}_{4C}$ & $\Tilde{\nu}_{harm}$  & $\chi$ & $\Tilde{\nu}_{harm}$ & $\chi$ & $\psi$ \\
         
$\left[ cm^{-1} \right]$ & $\left[ cm^{-1} \right]$ &$\left[ cm^{-1} \right]$ &$\left[ cm^{-1} \right]$&$\left[ cm^{-1} \right]$& &$\left[ cm^{-1} \right]$ & & \\
\hline

1786 & 3549 & 5267 & 6915 & 1798 & 0.0128 & 1764 & -0.0098 & -0.00217 \\
1788 & 3553 & 5282 & 6940 & 1799 & 0.0118 & 1753 & -0.0119 & -0.00229 \\
1790 & 3557 & 5285 & 6944 & 1801 & 0.012  & 1757 & -0.0115 & -0.00227 \\
1803 & 3588 & 5334 & 7006 & 1816 & 0.0119 & 1750 & -0.0157 & -0.00267 \\
1828 & 3634 & 5401 & 7107 & 1838 & 0.0111 & 1812 & -0.0077 & -0.0018  \\
1845 & 3671 & 5461 & 7183 & 1856 & 0.0109 & 1803 & -0.0127 & -0.00228 \\
1848 & 3670 & 5455 & 7191 & 1854 & 0.0101 & 1861 & -0.0016 & -0.00111 \\
1849 & 3679 & 5473 & 7212 & 1858 & 0.0098 & 1834 & -0.007  & -0.00162 \\
1858 & 3694 & 5492 & 7229 & 1867 & 0.0107 & 1839 & -0.0079 & -0.00179 \\
1861 & 3700 & 5505 & 7254 & 1869 & 0.0099 & 1851 & -0.0059 & -0.00151 \\
1861 & 3702 & 5507 & 7253 & 1870 & 0.0102 & 1843 & -0.0076 & -0.00171 \\
1801 & 3581 & 5322 & -    & -     &    -    &   -   & -        &    -      \\
1831 & 3638 & 5412 & -    & -     &    -    &     - &   -      &     -     \\
1835 & 3651 & 5430 & -    &  -    &    -    &  -    &   -      &   -       \\
1840 & 3654 & 5431 & -    & -     &    -    &      &  -       &     -     \\
1848 & 3681 & 5489 &  -    &  -    &    -    &  -    & -        &     -   \\ 

        \hline \hline
        \multicolumn{9}{c}{\textbf{Carbon atomic wires}} \\
        \hline
        \multicolumn{4}{c|}{Raman shift} & \multicolumn{2}{c|}{VPT2} & \multicolumn{3}{c}{VPT4} \\
        \hline
        
         $\Tilde{\nu}_{1\alpha}$& $\Tilde{\nu}_{2\alpha}$ & $\Tilde{\nu}_{3\alpha}$ & $\Tilde{\nu}_{4\alpha}$ & $\Tilde{\nu}_{harm}$  & $\chi$ & $\Tilde{\nu}_{harm}$ & $\chi$ & $\psi$ \\
         
$\left[ cm^{-1} \right]$ & $\left[ cm^{-1} \right]$ &$\left[ cm^{-1} \right]$ &$\left[ cm^{-1} \right]$&$\left[ cm^{-1} \right]$& &$\left[ cm^{-1} \right]$ & & \\
\hline
2098 & 4182 & 6226 & 8263 & 2125 & 0.0032 & 2146 & 0.0023   & -0.00008 \\
2098 & 4182 & 6226 & 8263 & 2125 & 0.0032 & 2146 & 0.0023   & -0.00008 \\
2124 & 4237 & 6334 & 8418 & 2099 & 0.0053 & 2160 & 0.0076   & 0.00023  \\
2172 & 4333 & 6481 & 8621 & 2068 & 0.0055 & 2081 & 0.0004 & -0.00048 \\

\end{tabular}

    \caption{Experimental Raman modes of confined carbyne (C mode) and carbon atomic wires (ECC or$\alpha$ mode), ideal harmonic frequency ($\Tilde{\nu}_{harm}$), and anharmonic parameters ($\chi$, $\psi$) for each carbyne-like system, calculated using Eq.~\ref{eq:linear_chi} for VPT2 and Eq.~\ref{eq:quadratic_chi_psi} for VPT4}
    \label{tab:all}\
\end{table}

\newpage
\section{Total anharmonic redshifts and look-up table}
In Fig.~\ref{fig:total_anh} we provide the total anharmonic redshift of the overtones of confined carbyne and carbon atomic wires relative to the corresponding multiple of fundamental BLA oscillation mode (C mode/ECC mode). The increase in the anharmonic overtone redshift with decreasing BLA oscillation follows linear trends and their parameters are provided in the Figure. The total anharmonic redshifts expected for ideal carbyne are indicated by the dashed line. We expect alternative carbyne-like materials to follow the same relations. Figure~\ref{fig:total_anh} may hence serve as a look-up table to confirm that a material is carbyne-like based on its BLA oscillation and overtone frequencies. 

\begin{figure}[h]
    \centering
    \includegraphics[width=\linewidth]{Images_SI/Total redshift supporting.pdf}
    \caption{ \textbf{Total anharmonic redshift} $\Delta\widetilde{\nu}_{tot,n} = n \cdot \widetilde{\nu}_{1} - \widetilde{\nu}_{n}$ as a function of the fundamental BLA oscillation mode frequency for confined carbyne and carbon atomic wires up to the 4$^{th}$ order. Linear trends and their equations are included. }
    \label{fig:total_anh}
\end{figure}

\newpage
\bibliography{references}